\documentclass{iopart}
\usepackage{amsmath,graphicx,bbm,iopams,mathrsfs,psfrag,pstricks,amssymb}
\newcommand{\ket}[1]{\vert #1 \rangle} \newcommand{\bra}[1]{\langle #1 \vert}

\newcommand{\bmsigma}{\boldsymbol \sigma}
\newcommand{\bmX}{\boldsymbol X} 
\def\M{{\hbox{\scriptsize M}}}
\begin{document}
\title{Bayesian estimation in homodyne interferometry}
\author{Stefano Olivares}
\address{ CNISM, UdR Milano Universit\`a, I-20133 Milano, Italy
\\ Dipartimento di Fisica, Universit\`a di Milano, I-20133 Milano, Italy}
\author{Matteo G. A. Paris}
\address{
Dipartimento di Fisica, Universit\`a di Milano, I-20133 Milano, Italy
\\CNISM, UdR Milano Universit\`a, I-20133 Milano, Italy
\\Institute for Scientific Interchange Foundation, I-10133 Torino, Italy}
\date{\today}
\begin{abstract}
We address phase-shift estimation by means of squeezed vacuum probe and
homodyne detection. We analyze Bayesian estimator, which is known to
asymptotically saturate the classical Cram\'er-Rao bound to the
variance, and discuss convergence looking at the a posteriori
distribution as the number of measurements increases. We also suggest
two feasible adaptive methods, acting on the squeezing parameter and/or the
homodyne local oscillator phase, which allow to optimize homodyne
detection and approach the ultimate bound to precision imposed by the 
quantum Cram\'er-Rao theorem. The performances of our two-step methods 
are investigated by means of Monte Carlo simulated experiments with a 
small number of homodyne data, thus giving a quantitative meaning to the 
notion of asymptotic optimality.
\end{abstract}
\maketitle
\section{Introduction}\label{s:intro}
Quantum phase measurements cannot be described by means of a proper
observable and different operational approaches have been introduced
over the years
\cite{PB:PRA89,NFM:PRL91,NFM:PRA92,NFM:PRA92b,cos93,rip94}.  On the
other hand, from a practical point of view, phase detection of quantum
fields is generally associated with interferometric devices, {\em i.e.},
detection schemes aimed at the estimation of phase by measuring field-
or intensity-based quantities with phase-dependent statistics
\cite{cav81,bon84,yur86,bur93,chi94,int95,nin97}.  The art of
interferomety, in turn, consists in answering to two question: a) How
can the unknown phase be effectively retrieved from the data sample? and
b) Which is the resulting precision?  The first point amounts to the
choice of an estimator, {\em i.e.}, a function from the data sample to
the set of possible values of the phase-shift. Among possible estimators
Bayes \cite{BIN} and maximum likelihood ones \cite{Lane,hraPRA96} play a
special role due to their asympotic ({\em i.e.}, for large number of
measurements) properties.  The second point may be properly addressed in
the framework of quantum estimation theory, which addresses the
inference of a physical quantity which is not directly accessible by
means of the measurement of a different observable, or a set of
observables, somehow related to the quantity of interest.  Quantum
estimation is a powerful tool to infer a single parameter, as well to a
set of parameters, up to the full reconstruction of the density matrix
of an unknown quantum state, with or without the use of prior
information \cite{HEL,LNP,ZH:PRA95}. Precision of any unbiased estimator
is bounded by the inverse Fisher information of the probability
distribution of the measurements outcomes, whereas the ultimate limit is
written in term of the inverse Quantum Fisher Information (QFI).
\par
In quantum optical systems, homodyne measurements of field quadratures
and Gaussian signals play a leading role. Indeed, measurement of
quadratures has been shown to achieve phase estimation for coherent
states with precision bounded by the (classical) Fisher information
\cite{hraPRA96}. This result have been further improved by looking for
the optimal state achieving the ultimate bound related to the QFI
\cite{braPRL94}.  Among the pure Gaussian states, squeezed vacuum has
been found to be the most sensitive state at fixed energy and homodyne
detection \cite{monPRA06}.  Furthermore, it has been shown that the same
signal allows optimal estimation of loss in bosonic channels
\cite{monPRL07} and of interaction parameters of single- and two-mode
bilinear bosonic Hamiltonians \cite{gaiba}.  Motivated by these results,
in this paper we address optimal phase estimation by using Gaussian
states, homodyne measurements and Bayesian estimation. We analyze the
behavior for increasing number of measurements and show that optimality
may be approached also with a limited number of runs upon using two-step
methods acting on the squeezed vacuum probe and/or on the homodyne
reference.  Moreover, we prove that, in principle, the performances of
double homodyne detection cannot beat the homodyne measurement ones,
thus validating the conclusions of \cite{monPRA06}.
\par
The paper is structured as follows. In Section \ref{s:estimation} we
briefly review local quantum estimation theory and the ultimate bounds
to precision in the phase-shift estimation by Gaussian states.  In
Section \ref{s:HDBI} homodyne and double homodyne statistics are
explicitly calculated for the phase shifted squeezed vacuum as input:
this leads us to conclude that performances of the double homodyne
detection cannot reach the limit imposed by QFI, while single homodyne
does.  Then, after describing our inference scheme, based on homodyne
detection and Bayesian inference, the asymptotic limit for large number
of collected data is studied in details, as well as the validity of the
Gaussian approximation. Since the performances of this kind of inference
protocol depend on the actual value of the (unknown) phase shift, we
suggest two feasible two-step adaptive methods \cite{wis,man}, the first
acting on the squeezing parameter, the other on the squeezing and local
oscillator phases, that allow always to reach the optimal estimation.
The results of simulated Monte Carlo experiments are reported in order
to check convergence also for small data sample and give a quantitative
meaning to the notion of asymptotic regime.  Section \ref{s:remarks}
summarizes our results and draws some concluding remarks.
\section{Estimation of a phase shift}\label{s:estimation}
Let us now consider a field mode undergoing a phase shift
described by the unitary operator $U(\phi)=\exp(-i\phi G)$, with
$G=a^\dagger a$,
$a$ and $a^\dagger$ being the annihilation and creation field operators,
respectively. Usually $\phi$ itself cannot be measured and a phase
estimation problem appears. In order to infer the value of $\phi$ some
phase-dependent observable $X$ is measured and an estimator
for $\phi$, {\em i.e.}, a function of the data sample $\{ x \}$ is used.
The aim of interferometry is to optimize the inference strategy by
minimizing the uncertainty. In general, the lower bound to the variance
${\rm Var}[\phi]$ of any unbiased estimator is given by the Cram\'er-Rao
theorem, which reads:
\begin{equation}
{\rm Var}[\phi] \ge [F(\phi)]^{-1},
\label{crb}
\end{equation}
where $F$ is the Fisher information:
\begin{equation}\label{CFI}
F(\phi) = \sum_x p(x|\phi)
\left[ \partial_\phi \log p(x|\phi) \right]^2,
\end{equation}
$p(x|\phi)$ being the conditional probability of obtaining the
outcome $x$ when the parameter has the value $\phi$.
Since the conditional probabilities are given by
$p(x|\phi) = {\rm Tr}(\varrho_\phi E_x)$, 
$\varrho_\phi = U(\phi)\varrho_0 U^\dag(\phi)$ being the
quantum state of the system (actually depending on the initial
preparation $\varrho_0$) and $E_x$ is the positive operator-valued measure
(POVM) describing the measurement, Eq.~(\ref{CFI}) rewrites as:
\begin{equation}\label{CFI2}
F(\phi) = {\rm Re} \, \sum_x
\frac{[{\rm Tr}(\varrho_\phi E_x \Lambda_\phi)]^2}
{{\rm Tr}(\varrho_\phi E_x)},
\end{equation}
where $\Lambda_\phi$ denotes the {\em symmetric logarithmic derivative}
(SLD) operator:
\begin{equation}\label{SLD}
\partial_\phi \varrho_\phi =
\frac12 (\Lambda_\phi \varrho_\phi + \varrho_\phi\Lambda_\phi).
\end{equation}
Upon using Schwartz inequality in the Hilbert space one easily shows
that the Fisher information in Eq. (\ref{CFI2}) is upper bounded
by the so-called quantum Fisher information QFI $H(\phi)$ 
\cite{braPRL94}, {\em i.e.}:
\begin{equation}\label{H:fish}
F(\phi) \le H(\phi) \equiv {\rm Tr}(\varrho_\phi \Lambda_\phi^2)\:.
\end{equation}
The above equation, togehter with the Cramer-Rao theorem sets the ultimate, 
measurement-independent, bound to precision of any unbiased estimator
involving quantum measurements.
\par
In order to calculate the SLD $\Lambda_\phi$, we first observe that
if $\varrho_0$, and, in turn, $\varrho_\phi$ are pure states,
then $\varrho_\phi = \varrho_\phi^2$ and $\partial_\phi \varrho_\phi^2 =
(\partial_\phi \varrho_\phi) \varrho_\phi +
\varrho_\phi (\partial_\phi \varrho_\phi)$, thus, by comparison with
Eq.~(\ref{SLD}), one finds $\Lambda_\phi = 2\, \partial_\phi \varrho_\phi$.
More in general, we can expand $\varrho_0$ in its eigenvector basis
$\{ \ket{\psi_n} \}$, {\em i.e.},
$\varrho_0 = \sum_n p_n \ket{\psi_n}\bra{\psi_n}$ (if $\varrho_0$ is a
pure state, then $p_n$ reduces to a Kronecker delta), to write:
\begin{equation}
\Lambda_\phi = \sum_{hk}\bra{\psi_h} \Lambda_\phi \ket{\psi_k}\,
\ket{\psi_h} \bra{\psi_k}\,. 
\end{equation}
Then, since
\begin{equation}
\partial_\phi \varrho_\phi =
i \sum_{hk} G_{hk} (p_h-p_k) \, \ket{\tilde\psi_h} \bra{\tilde\psi_k}
\end{equation}
with $G_{hk} = \bra{\psi_h} G \ket{\psi_k}$, where
$\ket{\tilde\psi_n} = U(\phi)\ket{\psi_n}$, we have:
\begin{equation}
\frac{\Lambda_\phi \varrho_\phi + \varrho_\phi\Lambda_\phi}{2} =
\frac12 
\sum_n p_n \left( \Lambda_\phi \, \ket{\tilde\psi_n} \bra{\tilde\psi_n} +
\ket{\tilde\psi_n} \bra{\tilde\psi_n} \Lambda_\phi \right)\,.
\end{equation}
By taking the matrix elements of both sides in Eq.~(\ref{SLD}) we obtain:
\begin{equation}
\bra{\tilde \psi_h} \Lambda_\phi \ket{\tilde \psi_k} =
\bra{\psi_h} \Lambda_0 \ket{\psi_k} = 
2 i G_{hk} \frac{p_h - p_k}{p_h + p_k},
\end{equation}
where $\Lambda_\phi \equiv U(\phi) \Lambda_0 U^{\dagger}(\phi)$.
As a consequence, $H(\phi) = {\rm Tr}(\varrho_\phi \Lambda_\phi^2) = 
{\rm Tr}(\varrho_0 \Lambda_0^2)$, {\em i.e.}, the QFI does not depend on
the value of the unknown shift $\phi$. The explicit evaluation of the 
QFI $H=H(\phi)=H(0)$ leads to:
\begin{align}
H &= 4 \sum_{ns} p_n \frac{(p_n - p_s)^2}{(p_n + p_s)^2} \, G_{ns}^2,
\label{H:start}
\end{align}
where we used $G_{ns}=G_{sn}$. The maximum is obtained for the probe 
excited in a pure state.  In this case, as described above,
$\Lambda_\phi = 2\,\partial_\phi \varrho_g$ and, by substitution
into Eq.~(\ref{H:fish}), we obtain $H = 4\Delta G^2$, 
{\em i.e.}, the QFI is proportional to the fluctuations of the Hamiltonian
$G$ and the ultimate bound of ${\rm Var}[\phi]$ becomes:
\begin{equation}\label{minVar}
{\rm Var}[\phi] = (4\Delta G^2)^{-1}.
\end{equation}
It is worth noticing that besides the number operator the above considerations 
hold for a general Hamiltonian generator $G$ \cite{gaiba}.
\par
Let us now come back to the problem of estimating $\phi$ by measurements 
on $\varrho_\phi$.
Our aim is to effectively estimate the phase shift at fixed energy upon 
optimizing the measurement over detection strategies and probe states
$\varrho_0$.
Of course, the ultimate precision is bounded by the quantum 
Cram\'er-Rao relation (\ref{minVar}), which depends on the probe state 
we employ. In turn, the first stage of the
optimization procedure is to find the best probe, which maximizes 
the QFI at fixed energy. We focus our attention
onto the set of {\em pure} states and, more precisely, on Gaussian
pure states, whose generic element is a squeezed-displaced vacuum state
given by:
\begin{equation}\label{GS}
\varrho_0 = D(\alpha) S(\xi) \ket{0}
\bra{0} S^{\dagger}(\xi) D^{\dagger}(\alpha),
\end{equation}
$D(\alpha) = \exp(\alpha a^{\dagger} - \alpha^* a)$ and
$S(\xi) = \exp(\frac12 \xi {a^{\dagger}}^2 - \frac12 \xi^* a^2)$,
$\alpha,\xi\in\mathbbm C$, being the displacement and squeezing operators,
respectively. In order to maximize the QFI 
we look for the state maximizing the
energy fluctuations at fixed probe energy ${\rm Tr}[\varrho_0 a^\dagger a] 
= \sinh^2 r + |\alpha|^2$,
\begin{align}
\Delta G^2 =& \frac12 \sinh^2 (2r) +
e^{2r} \left\{ {\rm Re}[\alpha]\cos\varphi +
{\rm Im}[\alpha]\sin\varphi \right\}^2 \nonumber\\
&-e^{-2r} \left\{ {\rm Re}[\alpha]\sin\varphi -
{\rm Im}[\alpha]\cos\varphi \right\}^2,
\end{align}
where we put $\xi = r e^{-2 i \varphi}$.
By using Lagrange multipliers one easily finds
$|\alpha|=0$: the maximum sensitivity is achieved when
all the available energy is used to squeezed the vacuum. 
Then we have: $\Delta G^2 =\frac12 \sinh^2 (2r)$ and 
thus
\begin{align}\label{minVar:Opt}
{\rm Var}_{\rm opt}[\phi] = [2 \sinh^2 (2r)]^{-1},
\end{align}
which represents the ultimate bound on precision of phase-shift estimation
posed by quantum mechanics (for Gaussian probes) \cite{monPRA06}.
Notice that Eq. (\ref{minVar:Opt}) does not depend
on the argument $\varphi$ of the complex squeezing parameter $\xi$:
without lack of generality we will assume $\varphi=\pi/2$.
In the next Section we will show how it is possible to attain the
ultimate precision by means of homodyne detection and Bayesian
inference.
\section{Phase-shift estimation via homodyne detection and
Bayesian inference}\label{s:HDBI}
\begin{figure}[h]
\includegraphics[width=0.6\textwidth]{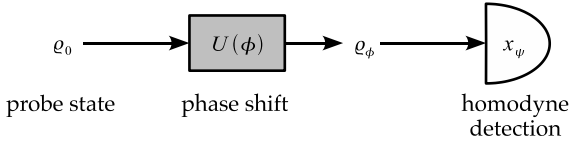}
\caption{\label{f:scheme} Scheme of phase estimation via
homodyne detection: an input state $\varrho_0$ undergoes a
phase shift $\phi$. The quadrature $x_\psi$ of the shifted state
$\varrho_\phi = U(\phi) \varrho_0 U^{\dagger}(\phi)$ is then measured
by means of homodyne detection.}
\end{figure}

We consider a general scheme (see Fig.~\ref{f:scheme}) where 
the probe state $\varrho_0$ undergoes a phase-shift and then the 
quadrature $x_\psi$ is measured by homodyne detection on the 
outgoing state, $\varrho_\phi$.
\\
The aim of our scheme is to infer the actual value $\phi$ of the phase
shift by processing the homodyne data.
In order to evaluate the homodyne probability distribution we
use the Wigner function formalism to describe our system.
The Gaussian Wigner function associated with the state (\ref{GS})
is (we put $\alpha = 0$ and $\varphi=\pi/2$):
\begin{equation}\label{wig0}
W_0(\bmX) = \frac{\exp[-\frac12 \bmX^T
\bmsigma_0^{-1} \bmX]}
{2 \pi \sqrt{{\rm Det}[\bmsigma_0]}},
\end{equation}
where $\bmsigma_0 = \frac14 {\rm Diag}(e^{-2r},e^{2r})$
is the $2\times 2$ covariance matrix.
After the phase shift (see Fig.~\ref{f:scheme}), the state $\varrho_\phi$
is still described by a Gaussian Wigner function $W_\phi(\bmX)$ of
the form (\ref{wig0}), but with covariance matrix $\bmsigma_\phi$ given by:
\begin{align}
[\bmsigma_\phi]_{11} &=\frac14 (e^{2r} \cos^2 \phi + e^{-2r} \sin^2 \phi),\\
[\bmsigma_\phi]_{22} &=\frac14 (e^{-2r} \cos^2 \phi + e^{2r} \sin^2 \phi),\\
[\bmsigma_\phi]_{12} &= [\bmsigma_\phi]_{21} =
\frac14 \sinh(2r)\,\sin(2\phi).
\end{align}
At this point the quadrature
$x_\psi = \frac12 (e^{-i\psi} a + e^{i\psi} a^{\dagger})$
is measured by means of homodyne detection on repeated preparation 
of the probe state, thus obtaining
a  data sample $\{ x \}$. Each outcome is distributed
according to the homodyne probability distribution, which can
be calculated starting from the Wigner function $W_\phi(\bmX)$
as follows:
\begin{equation}
p_\phi (x,\psi) = \int_{\mathbbm R} dy\, W_\phi(R_\psi \bmX),
\end{equation}
where $R_\psi$ is a rotation matrix
and $\bmX^{T} = (x,y)$. Since we put $\varphi=\pi/2$, we choose to
measure the quadrature with $\psi=0$. We have:
\begin{equation}\label{HD:prob}
p_{\rm H}(x|\phi) \equiv p_{\phi}(x,0) = 
\frac{1}{\sqrt{2\pi \Sigma_\phi^2}}
\exp\left( -\frac{x^2}{2 \Sigma_\phi^2}\right),
\end{equation}
where:
\begin{equation}\label{Sig:phi}
\Sigma_\phi^2 = \frac14
\left[ e^{-2r} \cos^2 \phi + e^{2r} \sin^2 \phi \right].
\end{equation}
The Fisher information of the distribution  
(\ref{HD:prob}) is given by:
\begin{align}
F_{\rm H}(\phi) &= \int_{\mathbbm R} dx\, p_{\rm H}(x|\phi)
\left[ \partial_\phi \log p_{\rm H}(x|\phi) \right]^2 
= \frac{\sinh^2(2r)\sin^2(2\phi)}{8(\Sigma_\phi^2)^2}.
\label{fisher:HD}
\end{align}
Remarkably, from Eq.~(\ref{fisher:HD}) we have that the Fisher 
information of homodyne distribution may be equal to the QFI 
upon the choice of a suitable squeezing of the probe state:
\begin{equation}
r = -\frac12 \log \tan \phi
\end{equation}
or, at fixed squeezing, for a specific value of the phase shift:
\begin{equation}\label{phi:H}
\phi_{\rm H} = \frac12 \hbox{arcos}\tanh 2 r\,.
\end{equation}
Correspondingly, the minumum variance ${\rm Var}_{\rm H}[\phi]$  
achievable by a suitable processing of homodyne data mat saturate 
$\forall \phi$ to the ultimate bound (\ref{minVar:Opt}).
\par
Before going to the Bayesian inference from of homodyne data,
we notice that if we use double-homodyne
detection we have no improvement in phase-shift estimation.
Double homodyne statistics is described by the coherent state POVM 
$\Pi_z = \pi^{-1} \ket{z}\bra{z}$, $\quad z\in{\mathbbm C}$; 
the probability distribution is thus given by:
$p_{\rm D}(z|\phi) = \pi^{-1} |\langle z | U (\phi) S(\xi) | 0 \rangle
|^2$ {\em i.e.}:
\begin{align}
p_{\rm D}(z|\phi) =
\frac{\exp\left\{ -|z|^2 -\tanh r\, {\rm Re}[z^2 e^{2i\phi}] \right\}}
{\pi \cosh r},
\end{align}
where we already set $\xi = -r$. The corresponding
Fisher information reads as follows:
\begin{align}\nonumber
F_{\rm D}(\phi) &= \int_{\mathbbm C}\!\! d^2z\, p_{\rm D}(z|\phi)
\left[ \partial_\phi \log p_{\rm D}(z|\phi) \right]^2 = 4\sinh^2 r,
\end{align}
that is $F_{\rm D}(\phi) \le F_{\rm H}(\phi)$, $\forall \phi$: 
the use of double homodyne detection does not bring any
improvement of the phase-shift measurement. This result agrees with
the conclusions of \cite{monPRA06}, where the author considered
double homodyne detection with squeezed vacuum as probe and an
auxiliary squeezed state in the other input port.
\par
We stress that  $p(x|\phi)$ allows us to infer the
probability of the homodyne outcome $x$ once the value of $\phi$ is 
assigned.  In our case, the value of $\phi$ is just the quantity we want 
to estimate and, in turn, we are interested in the conditional 
a posteriori probability distribution $p_\M(\phi|\{ x \})$ of 
$\phi$ {\em given} the the sample $\{x \}=\{x_1,...,x_M\}$ of homodyne
data. This can be obtained by means of Bayesian inference, as we will see 
in the following.
\subsection{Bayesian inference}
If $x$ is the random variable associated with the outcome
of the homodyne detection, then the Bayes' theorem states
that:
\begin{equation}\label{bayes}
p(x|\phi) p(\phi) = p(\phi|x) p(x)
\end{equation}
where $p(\cdot | \cdot)$ are the conditional probabilities, 
$p(\phi)=2/\pi$ is the prior assuming no a priori information, 
and $p(x)$ the overall probability to observe
$x$. In turn, upon inverting 
Eq.~(\ref{bayes}) we obtain the conditional a posteriori 
probability
$p(\phi|x)$ of $\phi$ given the outcome $x$. After $M$ independent
homodyne measurement the a posteriori probability is given by
\begin{equation}\label{bayes:1}
p_\M(\phi|\{ x \}) = \frac{1}{\cal N} \prod_{k=1}^{M} p(x_k|\phi),
\end{equation}
$\cal N$ being the normalization factor:
\begin{equation}
{\cal N} = \int_{0}^{\frac{\pi}{2}}\!\!\! d\phi \, p_\M(\phi|\{ x \}).
\end{equation}
If $M\gg 1$, then (\ref{bayes:1}) rewrites as:
\begin{equation}\label{bayes:2}
p_\M(\phi|\{x\}) \stackrel{M\gg1}{\simeq}
\frac{1}{\cal N} \prod_{x} p(x|\phi)^{M p(x|\phi^*)}\equiv p(\phi|M)
\end{equation}
where $\phi^*$ stands for the actual (unknown) value of the phase shift.
In order to write Eq.~(\ref{bayes:2}) we have used the law of large numbers
and written the number of occurrences of the outcome $x$ as $M p(x|\phi^*)$. 
In this limit probability (\ref{bayes:2}) can be explicitly calculated as
follows:
\begin{align}
p(\phi|M) \label{bayes:exp}
&= \frac{1}{\cal N} \exp\left\{
M \int\!\!dx\,p(x|\phi^*) \log p(x|\phi)
\right\} \\
&=\frac{1}{\cal N} \frac{1}{(2\pi \Sigma_\phi^2)^{M/2}}
\exp\left\{
-\frac{M \Sigma_{\phi^*}^2}{2 \Sigma_\phi^2}
\right\},\label{bayes:3}
\end{align}
where we used $\log\Pi_x \to \int dx$. We note that the
quantity $S(\phi|\phi^*)=-\sum_x p(x|\phi^*) \log p(x|\phi)$
in (\ref{bayes:exp}) may be regarded as the relative entropy
between the two distributions \cite{hraPRA96}.
\begin{figure}[tb]
\psfrag{pcond}{\footnotesize $p(\phi|M)$}
\psfrag{phi}{\footnotesize $\phi$}
\psfrag{sqfa}{\footnotesize $r=0.7$}
\psfrag{M1a}{\tiny $M=10$}
\psfrag{M2a}{\tiny $M=50$}
\psfrag{M3a}{\tiny $M=100$}
\includegraphics[width=0.41\textwidth]{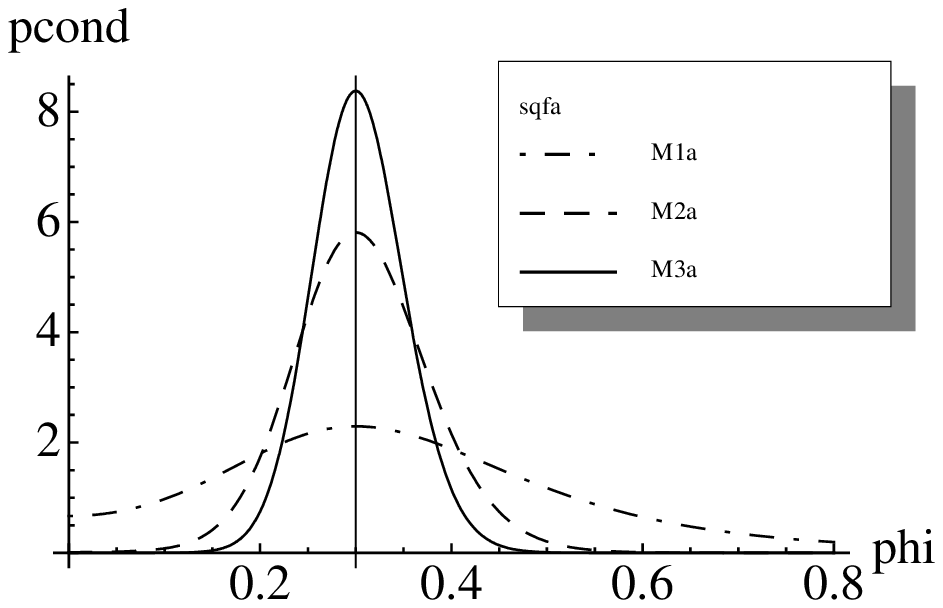}
\psfrag{sqfb}{\footnotesize $r=1.5$}
\psfrag{M1b}{\tiny $M=10$}
\psfrag{M2b}{\tiny $M=50$}
\psfrag{M3b}{\tiny $M=100$}
\includegraphics[width=0.41\textwidth]{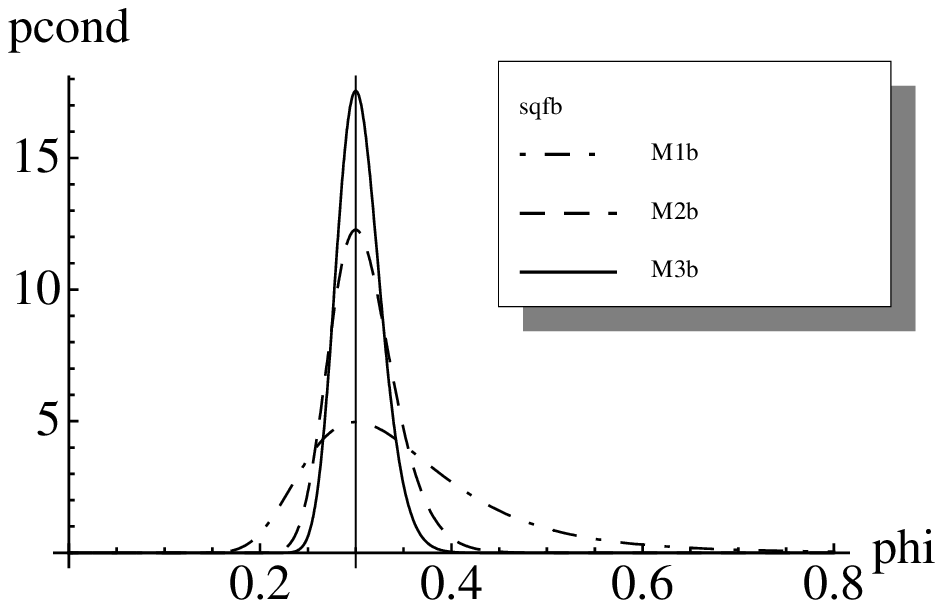}
\caption{\label{f:cond} A posteriori distribution $p(\phi|M)$
for different values the number of data $M$ ad squeezing parameter
$r$. The vertical line is the actual value of the phase shift $\phi^* = 0.3$.}
\end{figure}
In Fig.~\ref{f:cond} the a posteriori distribution $p(\phi|M)$ is plotted for 
different values of the involved parameters as a function of $\phi$. 
It is worth noting that because of the asymmetric form of the distribution,
a suitable estimator for the actual value $\phi^*$ of the phase shift is 
given by the maximum of the distribution
(its {\em mode}, ${\rm Mode}[\phi]$) and not to its mean:
$\overline{\phi}=
\int_{0}^{\frac{\pi}{2}}\!\!\!
d\phi\, \phi \, p(\phi|M)$.
This can be easily seen by differentiating $p(\phi|M)$ with 
respect to $\phi$:
\begin{equation}
\partial_\phi p(\phi|M) =
\frac{M p(\phi|M) F(\phi)}{8  \sin (2\phi)}
\left[ \cos(2\phi) - \cos(2\phi^*) \right],
\end{equation}
{\em i.e.}, $P(\phi|M)$ has a maximum at $\phi = \phi^*$.
However, as $M$ increases the mode and the mean become the same
and Eq.~(\ref{bayes:3}) can be approximated by a Gaussian
distribution \cite{brauJPA92} with mean $\phi^*$ and variance
$\Sigma_g^2$ given by:
\begin{align}
\Sigma_g^2 &= - \left[\frac{1}{p(\phi^*|M)}
 \left. \frac{d^2 p(\phi|M)}{d\phi^2} \right|_{\phi=\phi^*}
\right]^{-1}\label{Sigma:g}\\
&=\frac{1}{M}\left[
\sum_x \frac{1}{p(x|\phi^*)}
 \left. \frac{d^2 p(x|\phi)}{d\phi^2} \right|_{\phi=\phi^*}
\right]^{-1}
= \frac{1}{M F(\phi^*)},\label{BayesOptVar}
\end{align}
where we substituted Eq.~(\ref{bayes:exp}) into Eq.~(\ref{Sigma:g}) and
$F(\phi^*)$ is the Fisher information of the probaility 
distribution $p(x|\phi^*)$. 
The factor $M^{-1}$ follows from taking the data sample as a collection 
of $M$ mutually independent measurements, which, indeed, leads to an ensemble 
average over $M$ different copies of the system. Finally, we notice that,
as one may expect, the variance and, thus, the precision of the estimation 
depends on the true value $\phi^*$ itself.
\begin{figure}[tb]
\psfrag{Df}{\small$\Gamma$}
\psfrag{M}{\small $M$}
\psfrag{sqfa}{\footnotesize $r=0.6$}
\psfrag{sqfb}{\footnotesize $r=0.2$}
\psfrag{phit1}{\tiny $\phi^* = 0.3$}
\psfrag{phit2}{\tiny $\phi^* = 0.6$}
\psfrag{phit3}{\tiny $\phi^* = 0.9$}
\includegraphics[width=0.41\textwidth]{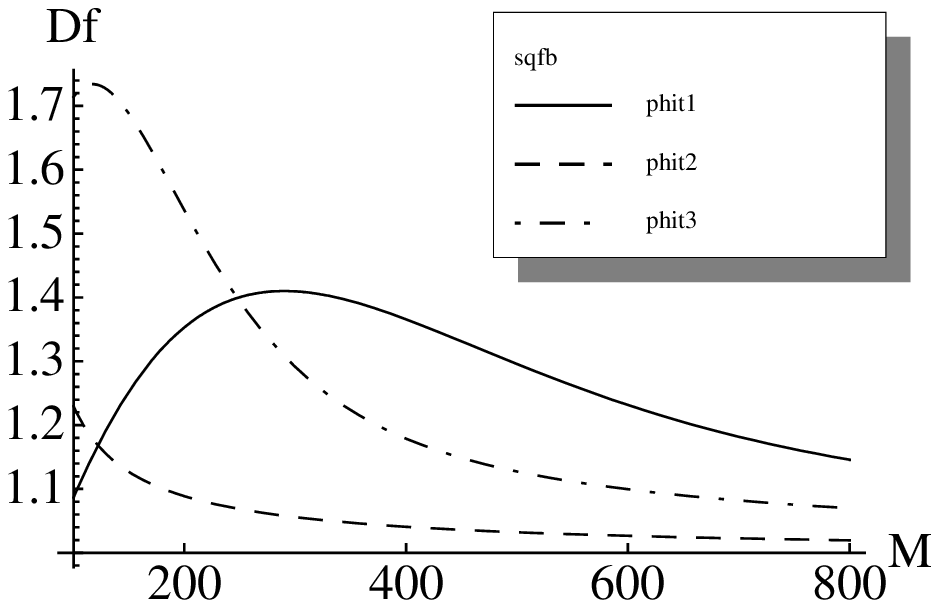}
\includegraphics[width=0.41\textwidth]{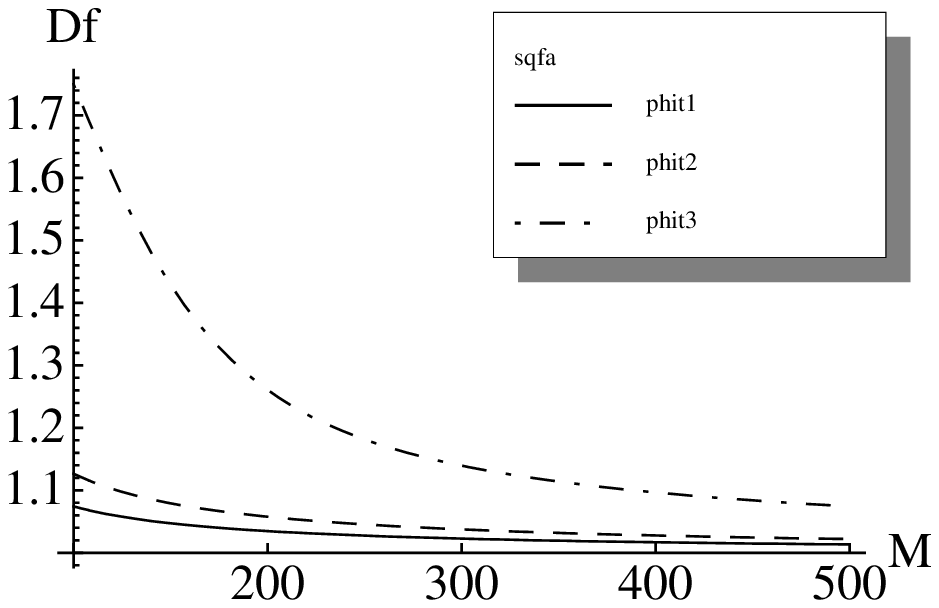}
\caption{\label{f:GammaSigma}  Plots of the ratio
$\Gamma=\Sigma_{\hbox{\scriptsize B}}^2/\Sigma_g^2$ for two values of
the squeezing parameter $r$ and different $\phi^*$. The range of
validity of Gaussian approximation strongly depends on the involved
parameters. In particular, the less is the difference between
$\phi^*$ and the optimal phase $\phi_{\rm H}$ given $r$
[see Eq.~(\ref{phi:H})], the larger is the range of validity of
this approximation (in the plots we have
$r=0.2 \rightarrow \phi_{\rm H}=0.59$
and $r=0.6 \rightarrow \phi_{\rm H}=0.29$).}
\end{figure} \\
Overall, the Bayes estimator is asymptotically
unbiased and efficient, {\em i.e.}, the variance ${\rm Var}[\phi]$
saturate the Cram\'er-Rao bound of Eq.~(\ref{crb}): this is
a consequence of the asymptotic normality of the a posteriori
distribution (Laplace-Bernstein-von Mises theorem) \cite{gill1,stat}.
However, two questions arises. The first concerns the range of validity
of  the Gaussian approximation, which depends on both $\phi^*$
and the squeezing parameter $r$. This aspect is illustrated in
Fig.~\ref{f:GammaSigma} where we plot the ratio
$\Gamma=\Sigma_{\hbox{\scriptsize B}}^2/\Sigma_g^2$,
$\Sigma_{\hbox{\scriptsize B}}^2$ being the variance of 
the asymptotic distribution $p(\phi|M)$. For fixed $r$, one finds that
the less is the difference between $\phi^*$ and the optimal phase
$\phi_{H}$ given $r$ [see Eq.~(\ref{phi:H})], the larger is the range
of validity of this approximation. On this observation is also based
the two-step adaptive method we will describe below.
The second question is whether the Bayes estimator may saturate
also the quantum Cram\'er-Rao bound, {\em i.e.}, whether the Fisher
information of $p(x|\phi)$ may be equal to the QFI, thus leading
to phase-shift estimation with precision at the ultimate quantum limit.
As what concerns this point we notice that, being the variance of Bayes 
estimator dependent on the true value of the phase shift, 
some kind of feedback should be unavoidably involved.
In the following we will describe two possible adaptive mechanisms,
acting on the squeezing parameter of the probe or on the
homodyne local oscillator and squeezing phase, respectively.
\subsection{Examples of two-step methods to achieve ultimate precision}
Adaptive methods for Bayesian estimation allow to always attain the ultimate 
bound on precision and have been investigated in the case of large
ensembles and qubit systems \cite{gill2,hay}. Here we propose two realistic 
and feasible setups exploiting the interferometric features of homodyne 
detection.
\par
\begin{figure}[tb]
\psfrag{Df}{\footnotesize $R(r)$}
\psfrag{r}{\footnotesize  $r$}
\includegraphics[width=0.6\textwidth]{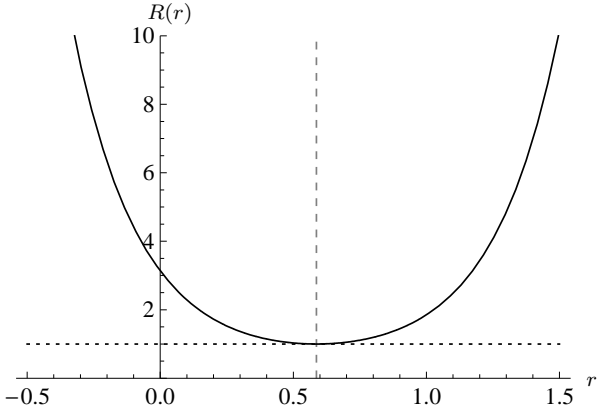}
\caption{\label{f:ratio} Plot of the ratio
$R(r) = M \Sigma_g^2(r)/{\rm Var}_{\rm opt}[\phi^*]$ (see text) 
as a function of $r$ and for $\phi^* = 0.3$. The vertical dashed 
line indicates $r_{\rm opt}$.}
\end{figure}

The first scheme is based on the fact that the variance $\Sigma_g^2(r)$
may achieve the optimal value
$M^{-1}{\rm Var}_{\rm opt}[\phi^*]$ of Eq.~(\ref{minVar:Opt})
employing a squeezed vacuum probe with parameter 
$r_{\rm opt} = -\frac12 \log \tan \phi^*$.
Of course, setting $r=r_{\rm opt}$ requires the knowledge of the actual
(unknown) value of the phase shift. However, one
may obtain a rough estimate of $\phi^*$ upon building
the distribution $p(\phi|M')$ with a fraction of the 
$M$, taking its maximum (${\rm Mode}[\phi]$) and then modify the probe
state,  tuning its squeezing to $r_{\rm opt}$. In Fig.~\ref{f:ratio} 
we show the ratio $R(r) = M \Sigma_g^2(r)/{\rm Var}_{\rm opt}[\phi^*]$ 
for the case $\phi^*=0.3$: the smooth behavior of
$R(r)$ ensures the convergence of the above mechanism.
Tuning the squeezing parameter, however, could be a challenging task.
On the other hand, also when $r$ and, thus, the energy are fixed,
it is possible to achieve the optimal variance by tuning the
squeezing phase $\varphi$ of the probe state or the phase $\psi$
of the homodyne quadrature. In fact, previously we set
$\varphi=\pi/2$ and $\psi=0$; if, on the contrary, we assign to
these phases the generic values $\varphi$ and $\psi$, then
we should simply apply the following change of variable
in all the previous equations:
$\phi \to \phi + \left(\varphi-\psi -\frac{\pi}{2} \right)$, 
that is a translation of $\phi$ by the amount
$\varphi-\psi -\frac{\pi}{2}$.
Since the optimal angle $\phi_{\rm H}$ at fixed $r$ is given by
Eq.~(\ref{phi:H}), optimality is always achieved by choosing:
\begin{equation}\label{Adp}
\varphi-\psi = \phi_{\rm H}-\phi^* + \frac{\pi}{2}.
\end{equation}
As described above, we may obtain a rough estimate of $\phi^*$
by taking the maximum (${\rm Mode}[\phi]$) of $p(\phi | M)$, and,
thus, we can tune the quantity $\varphi-\psi$ by means of adaptive
control on the homodyne detection and/or the probe state, whose energy
does not depends on $\varphi$. 
\par
\begin{figure}[tb]
\psfrag{fiR}{\small ${\cal A}$}
\psfrag{VarR}{\small ${\cal V}$}
\psfrag{M}{\footnotesize $M$}
\includegraphics[width=0.35\textwidth]{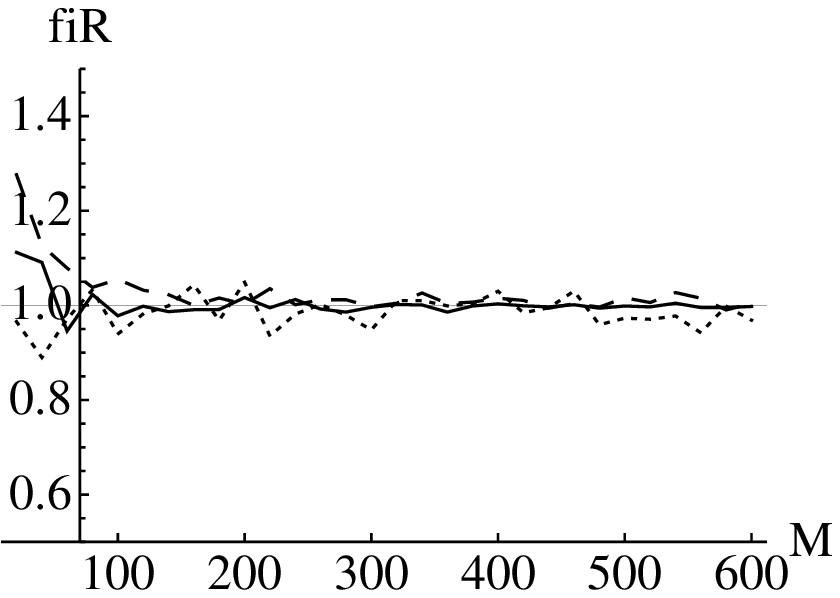} \qquad
\includegraphics[width=0.35\textwidth]{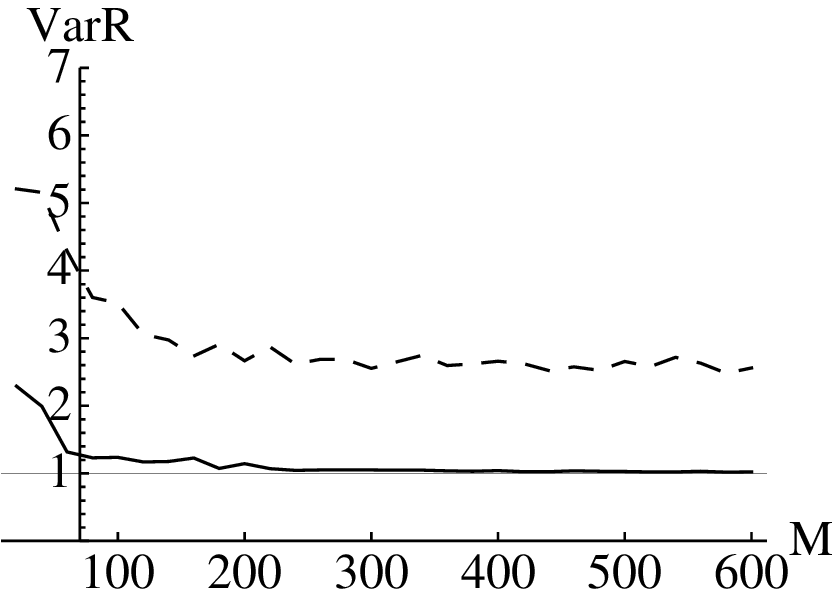} \\
\includegraphics[width=0.35\textwidth]{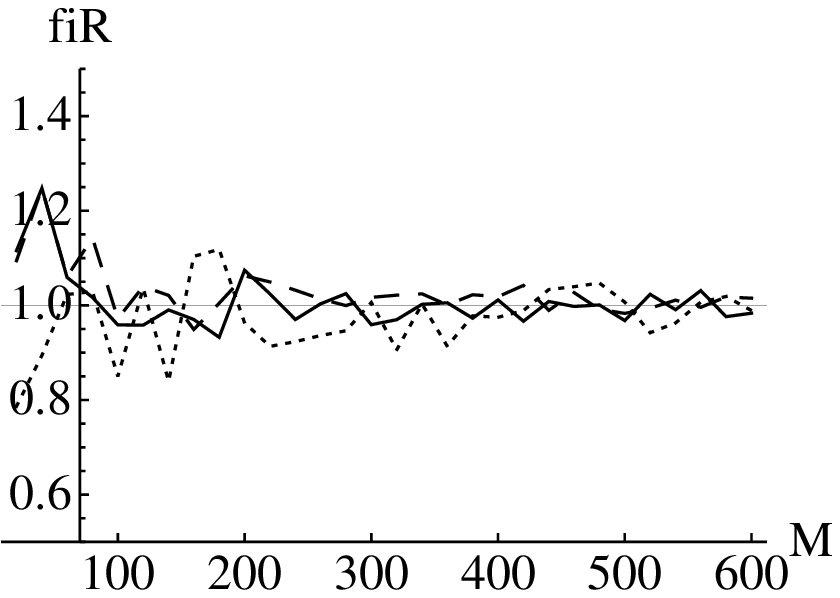}\qquad
\includegraphics[width=0.35\textwidth]{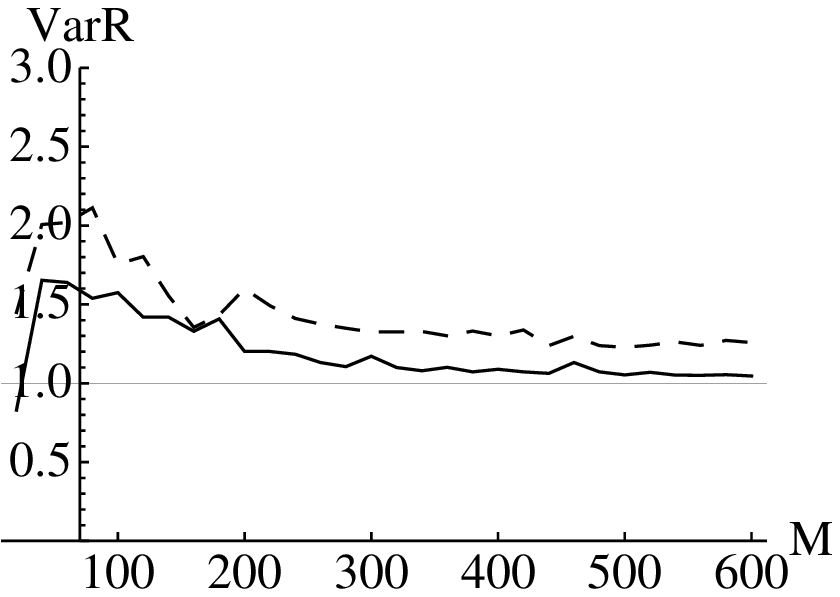}
\caption{\label{f:MC:exp} Bayesian estimation of the phase-shift from
Monte Carlo simulated homodyne measurements: on the left the ratio
${\cal A}=\overline{\phi}/\phi^*$ and on the right ${\cal V}=\sqrt{{\rm
Var}[\phi]/{\rm Var}_{\rm opt}[\phi]}$ (right).  The solid lines denote
results obtained with the adaptive method acting on the homodyne local
oscillator and squeezing phases; the dashed lines are obtained without
the adaptive method.  We set $r=0.6$ and $\phi^*=0.7$.  In the case of
the adaptive method, we used $\lfloor 3 \sqrt{M} \rfloor$ of the $M$
data to argue the phase-shift rough estimate (dotted line), then the
left homodyne data are processed to assess $\overline{\phi}$ and ${\rm
Var}[\phi]$. In both the experiments we use the same total number $M$ of
data. Lower panel: the same as in the top panel for $r=0.3$.}
\end{figure}
In order to confirm convergence also for small data sample, we performed
a set of Monte Carlo simulated experiments with the latter adaptive
scheme. The results are shown in Fig.~\ref{f:MC:exp} for $r=0.6$ and
$\phi^*=0.7$ (upper panel) and with reduced energy, $r=0.3$ 
(lower panel).  In the experiment without adaptive method the whole
sample of $M$ homodyne data, obtained as described in the first part of
this section, is used to estimate $\overline{\phi}$ and ${\rm
Var}[\phi]$ (dashed lines in Fig.~\ref{f:MC:exp}). With the adaptive
scheme (solid lines), $N_{\rm r} = \lfloor 3 \sqrt{M}\rfloor$ of the $M$
data sample are used to argue the phase-shift rough estimate, then the
phase difference $\varphi-\psi$ is tuned according to Eq.~(\ref{Adp})
and the left homodyne data are processed to assess $\overline{\phi}$ and
${\rm Var}[\phi]$.  Each point in Fig.~\ref{f:MC:exp} corresponds to the
average over 20 repetitions.
Of course, the effectiveness of the adaptive method depends on the
value of the rough estimate: in this view, an increasing number of the
outcomes devoted to the rough estimation, as the data sample becomes
larger, allows the reduction of the ${\rm Var}[\phi]$ fluctuations, as
one may verify, for example, by using a fixed value for $N_{\rm r}$.
It is worth to note that in our simulations the rough estimate is
obtained as ${\rm Mode}[\phi]$, whereas the mean $\overline{\phi}$
is used for the final results: this is justified for the small
$N_{\rm r}$ considered in the rough estimate and the larger
number of the final estimation
(the error introduced by this choice does not sensitively affect our
results, as we verified also assessing the Pearson skewness coefficient
$|\overline{\phi} - {\rm Mode}[\phi]|/\sqrt{{\rm Var}[\phi]}$).
\section{Conclusions}\label{s:remarks}
In this paper we have shown how Bayesian inference techniques represent
useful tools for phase estimation. Our analysis is based on homodyne detection
with squeezed vacuum as a probe state, and Bayesian post-processing to
infer the phase shift. In the asymptotic limit of a large number of
measurements, our scheme saturates the Cram\'er-Rao bound to precision,
{\em i.e.}, the variance of the phase shift achieves the lower bound imposed
by the inverse Fisher information.  Moreover, we have shown that
optimality may be approached also with a limited number of measurements
by means of two-step methods acting on the squeezed vacuum probe and/or
on the homodyne reference. These have been investigated by means of
Monte Carlo simulated experiments, which show excellent results also in
the case of small data samples. Our results, together with the recent
advances in homodyne detection \cite{par07} lead us to conclude that the
estimation protocol described in our paper may be suitable for
experimental investigation, opening the way to information technology
based on Gaussian states and phase encoding.
\section*{Acknowledgments}\label{s:ackn}
The authors thank A.~Monras and M.~G.~Genoni for useful discussions.
MGAP thanks Luca Pezze and Augusto Smerzi for useful discussions in the
early stage of this work.
This work has been partially supported by the CNR-CNISM convention.
This article was completed at a time of drastic cuts to research budgets
imposed by the Italian government; as a result research is becoming
increasingly difficult in Italian universities and may in the near
future be brought to a complete halt.

\end{document}